\documentclass[reprint,aps,superscriptaddress,amsmath,amssymb,article]{revtex4-2}
\usepackage{bm}        
\usepackage{amssymb}   
\usepackage{color}
\usepackage{units}
\usepackage{amsmath}
\usepackage{natbib}
\usepackage{esint}
\usepackage{chemformula}
\usepackage{ulem}
\usepackage[english]{babel}
\usepackage[colorlinks=true, citecolor=blue]{hyperref}
\usepackage{xcolor}
\usepackage{chemformula} 
\usepackage[T1]{fontenc} 
\usepackage{tikz}
\usepackage{parskip}
\usepackage{multirow}
\usepackage{graphicx}
\usepackage{mathtools} 
\usepackage{soul}
\usepackage{amsmath}
\DeclareUnicodeCharacter{2009}{\,}

\newcommand{\SNBose}{\affiliation{
		Department of Condensed Matter and Materials Physics, SN Bose National Centre for Basic Sciences, JD Block Sector III, Salt Lake, Kolkata - 700106}}
  
\newcommand{\IACS}{\affiliation{School of Physical Sciences, Indian Association for the Cultivation of Science, Jadavpur, Kolkata 700032, India}}

\begin{document}

\title{
Centrosymmetric-noncentrosymmetric Structural Phase Transition in Quasi one-dimensional compound, (TaSe$_4$)$_3$I}
\author{Arnab Bera}
\altaffiliation{These authors contributed equally to this work}
\IACS

\author{Samir Rom}
\altaffiliation{These authors contributed equally to this work}
\SNBose

\author{Suman Kalyan Pradhan}
\IACS

\author{Satyabrata Bera}
\IACS

\author{Sk Kalimuddin}
\IACS

\author{Tanusri Saha-Dasgupta}
\email{t.sahadasgupta@gmail.com}
\SNBose

\author{Mintu Mondal}
\email{mintumondal.iacs@gmail.com}
\IACS

\begin{abstract}
(TaSe$_4$)$_3$I, a compound belonging to the family of quasi-one-dimensional transition-metal tetrachalcogenides, has drawn significant attention due to a recent report on possible coexistence of two antagonistic phenomena,  superconductivity and magnetism below 2.5~K (Bera et. al, arXiv:2111.14525). Here, we report a structural phase transition of the trimerized phase at temperature, $T~\simeq$~145~K using Raman scattering, specific heat, and electrical transport measurements. The temperature-dependent single-crystal X-ray diffraction experiments establish the phase transition from a high-temperature centrosymmetric to a low-temperature noncentrosymmetric structure, belonging to the same tetragonal crystal family. The first-principle calculation  finds the aforementioned inversion symmetry-breaking structural transition to be driven by the hybridization energy gain due to the off-centric movement of the Ta atoms, which wins over the elastic energy loss.

\end{abstract}
\maketitle


\section{Introduction}

Symmetry plays a crucial role in dictating the properties of materials\cite{Zak1985,Gross1996,BookSym2007,Fu2007}. In particular, underlying crystal symmetry in reduced dimensions can have an important influence on their physical properties. 
Phase transitions accompanied by symmetry breaking in low-dimensional materials can endow fascinating phenomena, \cite{Armitage2018,Zhang2010,PhysRevB.93.134512,Shi2016,Du2021,Wang2016,Berger2018,Bode2007,Vescoli,Pei2021,NbSe2006,Fahad2023}  for instance, ferroelectricity, associated with the loss of inversion symmetry \cite{Anderson1965,Shi2013}, leads to spontaneous charge polarization. Similarly, in low-dimensional materials with broken inversion symmetry, spin-orbit coupling can result in novel spin textures and topological phases\cite{Sato2017,Matano2016, Lu2015,Liu2016}.

The family of transition metal tetrachalcogenides with generic formula (MX$_4$)$_n$Y (M = Nb, Ta; X = S, Se; Y = halogens like Br, I and $n$ = 2, 3, 10/3) serve as a model $quasi$-1D system for studying wide variety of phases and phase transitions \cite{Monceau1982,Gressier1984,Gressier1984a, Roucau_1984,Izumi1984,Meerschaut1984,Zwick1985,PhysRevB.37.1024,Taguchi1986,Sekine1987,Sekine1988,Ramirez,Balandin2022,PhysRevB.94.104113,Kusz2010,An2020}. These quasi-1D materials are composed of "M" atoms, surrounded by antiprisms of "X" atoms and arranged in 1D chain geometry, with "Y" ions between the chains giving cohesion. They are prone to Peierls instability and the filling of the metal $d$$_z$$^2$ band, given by $(n-1)/2n$, governs the 
nature of charge density wave (CDW) fluctuations driven by the correlation effect\cite{Gressier1984a}. One member of the above family, (TaSe$_4$)$_2$I is a chiral quasi-1D compound without an inversion center and has recently been claimed to be a topological Weyl semimetal, whose Weyl nodes are coupled with CDW modes, giving rise to an axion insulator \cite{Gooth2019,Li2021,Shi2021,Zhang2020,Mu2021,BalandinNoise2023}. This proposal opens up a novel avenue for investigating the interplay of strongly correlated CDW condensate and topology of electronic state,  which may lead to the emergence of unusual quantum phenomena \cite{Gooth2019,Kim2023}.

 Another member of this family, (TaSe$_4$)$_3$I, has caught attention in recent times due to a provocating report on the possible coexistence of ferromagnetism
and superconductivity below 2.5~K \cite{arnab2021nTSI}. The first experimental report by C Roucau, et al in the 1980s\cite{Roucau_1984} on (TaSe$_4$)$_3$I, showed that this compound is stabilized in crystal structure with space group, $P4/mnc$ at room temperature, and exhibits weakly metallic behavior \cite{Roucau_1984}. Despite being synthesized decades ago, the study on (TaSe$_4$)$_3$I is found to be rather limited. Previous studies assumed that (TaSe$_4$)$_3$I behaves similarly to (NbSe$_4$)$_3$I due to their identical crystal structures \cite{Gressier1984,Roucau_1984,Gressier1984a}. However, the recent findings\cite{arnab2021nTSI} demand separate attention to this compound, to understand the proposed
exotic behavior. Theoretical calculations on (TaSe$_4$)$_3$I \cite{Ramirez} using the crystal structure derived from (NbSe$_4$)$_3$I by replacing Nb with Ta, suggest the presence of a van-Hove singularity in the low-energy electronic structure of the material, which may contribute to its unique electronic properties. However, the origin of ferromagnetism co-existing with superconductivity remains elusive. Therefore, an understanding of the symmetry of the LT crystal structure is necessary as the reported exotic properties are observed at low temperatures.

The present study aims to investigate the inversion symmetry-breaking structural phase transition observed in a quasi-1D compound, (TaSe$_4$)$_3$I through a combination of experimental and theoretical approaches. The electronic transport, specific heat, and Raman scattering experiments reveal a clear phase transition at around $T_S \sim$145~K. The temperature-dependent single-crystal X-ray diffraction (SXRD) studies suggest that the phase transition is caused by a temperature-induced distortion of Ta-chains that breaks the inversion symmetry at low temperatures, leading to a structural phase transition from a  high-temperature (HT) centrosymmetric ($P4/mnc$) structure to a low-temperature (LT) noncentrosymmetric structure (P$\Bar{4}$2$_1$c).  We employ first principle calculation together with ab-intio derived tight binding formulation to  elucidate the microscopic origin of this inversion symmetry-breaking transition.
The transition is found to arise from the gain in hybridization energy by the off-centric movement of Ta atoms, in the chain-like crystal structure, overpowering the elastic energy loss, akin to the mechanism of lattice-driven ferroelectricity. Our findings provide valuable insight into the unique symmetry-breaking structural transition in a quasi-1D system and will motivate further investigation of the impact of this symmetry-breaking structural transition on its physical properties.

\section{EXPERIMENTAL AND CALCULATION DETAILS}

(TaSe$_4$)$_3$I single crystals were grown using the chemical vapor transport (CVT) method with iodine as the transport agent, following the procedure described in detail in earlier work \cite{arnab2021nTSI}. Temperature-dependent single crystal X-ray diffraction (SXRD) experiments were carried out on a Bruker D8 VENTURE Microfocus diffractometer equipped with a PHOTON II Detector, using Mo K$\alpha$ radiation ($\lambda$ = 0.71073 $^{\circ}$). The raw SXRD data were integrated and corrected for Lorentz and polarization effects using the Bruker APEX III program suite. Absorption corrections were performed using SADABS. The space groups were assigned by considering the systematic absences determined by XPREP and analyzing the metric symmetry. The space groups were further verified for additional symmetry using PLATON \cite{Spek2003,Spek2009}. The crystal structure was solved by direct methods and refined against all data in the reported 2$\theta$ ranges by full-matrix least-squares refinement on $F_2$ using the SHELXL \cite{Sheldrick2008} program suite and the OLEX2 \cite{Dolomanov2009} interface.

Electronic transport measurements were carried out in standard four-probe technique on a few ribbon-like oriented single crystals using a "Keithley 2450 source meter" in Cryogenics 16 Tesla measurement system. For the heat capacity measurement, a pallet was prepared by lightly pressing a bunch of single crystals in a palletizer. Specific heat measurement of as prepared pallet were done in the "Physical Properties Measurement System (PPMS)" by Quantum Design. To uncover the nature of phase transition, Raman scattering experiments were carried out on a single crystal sample of (TaSe$_4$)$_3$I across the phase transition temperature, $T_S$~$\sim$145~K using a 532 nm laser excitation in Horiba T64000 Raman spectrometer (with spot size $\sim$ 1~$\mu$m). 
 
 The first-principles calculations were carried out with three different choices of basis sets, (a) plane-wave basis as implemented in Vienna Abinitio Simulation Package (VASP)\cite{vasp-1, vasp-2}, (b) full potential linearized augmented plane wave (LAPW) as implemented in the Wien2k code\cite{wien2k}, and (c) linear and N-th order muffin-tin orbital basis \cite{lmto, nmto}. The consistency of the results in three different basis
 sets has been confirmed in terms of band-structures and density of states plots. The exchange-correlation functional was chosen as Perdew-Burke-Ernzerhof (PBE) implementation of generalized gradient approximation. 8 $\times$ 8 $\times$ 4 Monkhorst-Pack k-point 
 mesh was found to give good convergence of the computed ground-state properties. Plane-wave cut-off of 600 eV was used in plane-wave calculations with projected augmented wave potentials. For LAPW calculations, the criterion used was muffin-tin radius multiplied by K$_{max}$ for the plane wave yielding a value of 7.0.   The muffin tin radii for the LMTO calculations were chosen to be 3.06, 3.05, 3.32, and 2.50-2.60 {\AA} for Ta (1), Ta (2), I, Se, respectively in the case of HT (300~K) structure, and 3.06, 3.06, 3.05, 3.30 and 2.55-2.70 {\AA} for Ta (1), Ta (2), Ta (3), I and Se respectively, for the LT (100~K) structure. The plane wave, LAPW, and LMTO calculations have been used to calculate the electronic structure both at HT and LT phases, to check the robustness of the semi-metallic solution at HT and LT phases.

The construction of low energy Hamiltonian in first principles derived Wannier function basis, for the purpose of evaluation of kinetic energy gain in LT phase, was achieved through NMTO-downfolding technique starting from a full DFT band structure. The NMTO calculations have been carried out with the potential obtained from self-consistent LMTO calculation. The real space representation of the NMTO-downfolded Hamiltonian, 
$H_{TB} = \sum_{ij} t^{mm'}_{ij} (C^\dagger_{i,m} C_{j,m'} + h.c)$ in the Wannier function basis gives the estimates of various hopping integrals ($t$) where $m$ and $m'$ are the non-downfolded orbitals at sites $i$ and $j$, and $C^\dagger_{i,m} (C_{j,m'})$ are electron creation (annihilation) operators.

\section{Results and discussion}

\begin{figure*}
\begin{center}

\includegraphics[width=2\columnwidth]{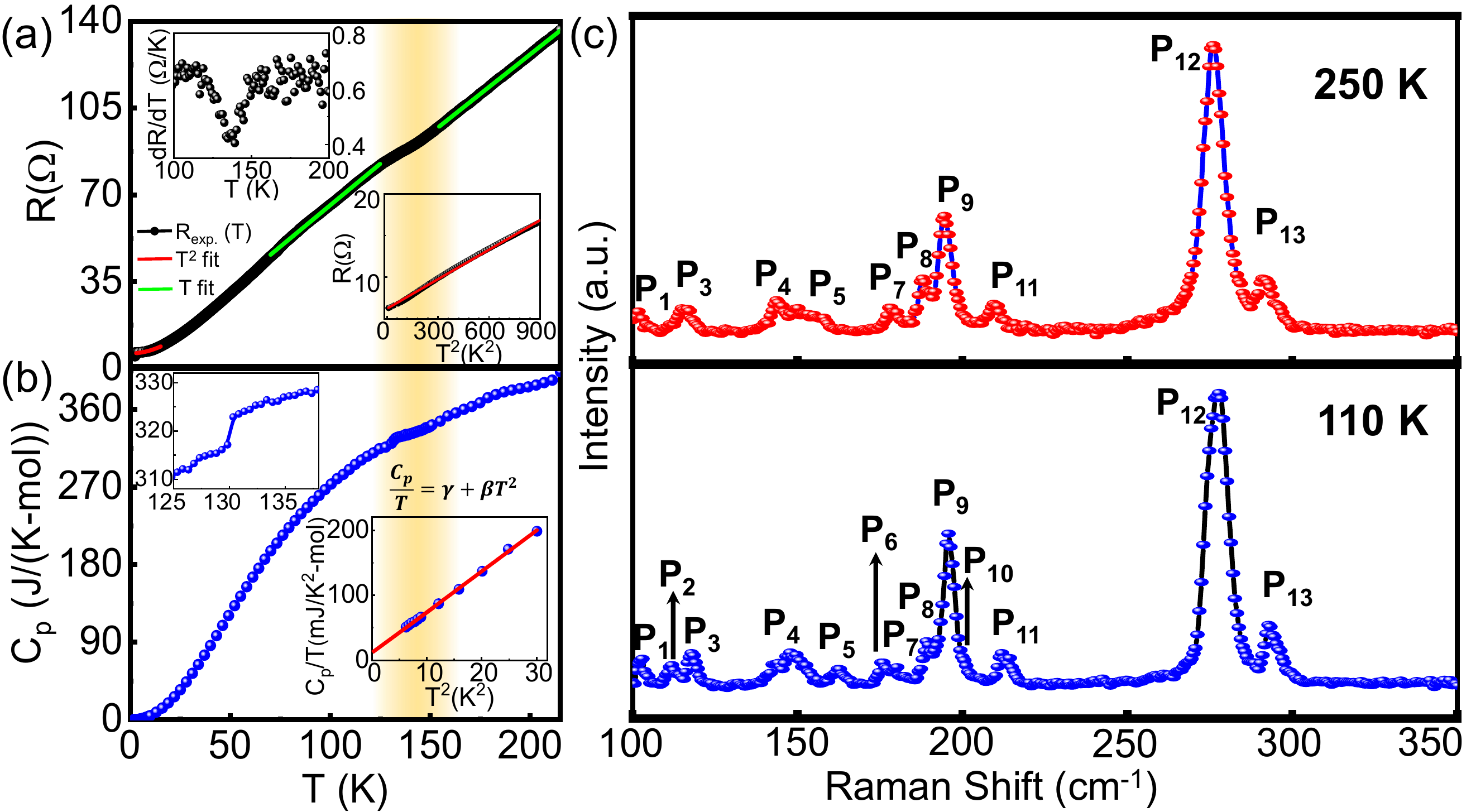}
		\caption {\textbf{Signature of phase transition in (TaSe$_4$)$_3$I :} (a) Temperature-dependent resistance measurement of wire like samples of (TaSe$_4$)$_3$I down to $T$ $\sim$ 2 K. The upper inset  (d$R$/d$T$ versus $T$ plot)  shows the phase transition at around 145~K. Bottom inset: $R$ vs. $T^2$ plot in the 2.5–30 K range. The red solid line is a linear fit to the experimental data. (b) The specific heat data shows a sharp jump (zoom in upper inset) which is a clear sign of a phase transition (the region presented in light yellow color), at around, $T_S \sim$ 145~K. Lower inset: C$_p$/T vs T$^2$, exhibiting linear behavior at low-temperature region.  (c) Raman spectra, performed at sample temperature, $T$ = 250 K and 110 K. Arrows indicate the  Raman-active modes P$_2$, P$_6$, and P$_{10}$ that appear in the low-temperature sample below $T_S$.} 
        \label{fig1 raman and cp}
	\end{center}
\end{figure*}

\subsection{Phase transition at $T_S \simeq 145~K$}

Figure~\ref{fig1 raman and cp}(a)  shows the measured temperature dependence of resistance ($R (T)$), and it decreases almost linearly with decreasing temperature ($T$). The obtained residual resistance ratio (RRR), $\frac{R_{300K}}{R_{2K}} = 30$, suggests that the sample is semi-metallic. Additionally, the $R$($T$) curve shows a clear slope change at around 145~K (denoted by $T_s$), which is also evident in the d$R$/d$T$ vs $T$ plot (see upper inset) implying a phase transition. To get further insight into transport properties, we analyzed the temperature dependence of longitudinal resistance data using Matthiessen’s rule\cite{Volkenshtein1973,Raquet2002}.

\begin{equation}
    R_{xx}(T)=R_0+R_{e-p}(T)+R_{e-e}(T)
    \label{Matthiessen}
\end{equation}
Where, R$_0$ is the residual resistivity arising from temperature-independent elastic scattering from static defects, R$_{e-p}$ represents the electron-phonon scattering contribution, which is proportional to $T^2$, while R$_{e-e}$ is the electron-electron scattering contribution proportional to $T$.
The temperature dependence of $R (T)$ shows linear behavior in the temperature range 70~K-200~K, below and  above the transition $T_S\sim 145~K$, suggestive of the dominance of the strong electron-phonon scattering mechanism  in this temperature range. Below 15~K, $R$ is found to vary quadratically with $T$ (see lower inset in Figure~\ref{fig1 raman and cp}(a)), indicating a strong electron-electron scattering mechanism that dominates the transport properties in low $T$ regime \cite{wilson2011theory}. No thermal hysteresis and significant change in $T_S$ was observed during repeated cooling and heating cycle under a high magnetic field ($H$), implying the absence of any magnetic contribution. 

To get further insight into this transition, we carried out temperature-dependent specific heat, $C_{\text{p}}$($T$) measurements of a pallet prepared using multiple (ribbon-like) single crystals  in the temperature range 2.8-250~K. The aforementioned phase transition is manifested as a sharp change or step-like anomaly in $C_{\text{p}}$ at $T$ $\sim$ 140~K (see Figure~\ref{fig1 raman and cp}(b)) \cite{Kratochvilova2017}. It is to be noted, there is a slight difference in transition temperatures obtained from transport and specific heat measurements arising due to the use of samples with different geometry. To estimate the electronic contribution of specific heat (Sommerfeld coefficient) the C$_p(T)$ was analyzed using the equation, $\gamma T+\beta T^3$ in  the temperature range, 2.5 to 6 K. Inset of Figure~\ref{fig1 raman and cp}(b) shows the C$_p$ versus $T^2$ plot for better clarity. The $\gamma$T represents the electronic contribution and the $\beta$$T^3$ is the lattice contribution to the specific heat. The obtained Sommerfeld coefficient, $\gamma$ = 11.80 mJ/K$^2$-mol and $\beta$ = 6.25 mJ/K$^4$-mol. The large value of $\gamma$ suggests the presence of a significant electronic correlation in the sample\cite{Shulenburger2008}.

 In Figure~\ref{fig1 raman and cp}(c), we present the Raman spectra recorded at temperatures of 110~K and 250~K. At 250~K, we identified ten Raman-active modes, whereas the spectra at 110~K revealed thirteen Raman-active modes, as shown in Figure~\ref{fig1 raman and cp}(c). The emergence of three new Raman-active vibration modes (P$_2$, P$_6$, and P$_{10}$) at lower temperatures suggests that the crystal symmetry of (TaSe$_4$)$_3$I no longer remained identical below the transition\cite{Bera2023Raman,Sekine1988,Dominko2016,Gao2018,Zhang2016}. The resistivity, and specific heat data corroborated with Raman data thus reveal a structural phase transition, the detailed nature of which we further investigated using temperature-dependent single-crystal X-ray diffraction(XRD), as presented below. 

\begin{figure*}
\begin{center}
\includegraphics[width=2\columnwidth]{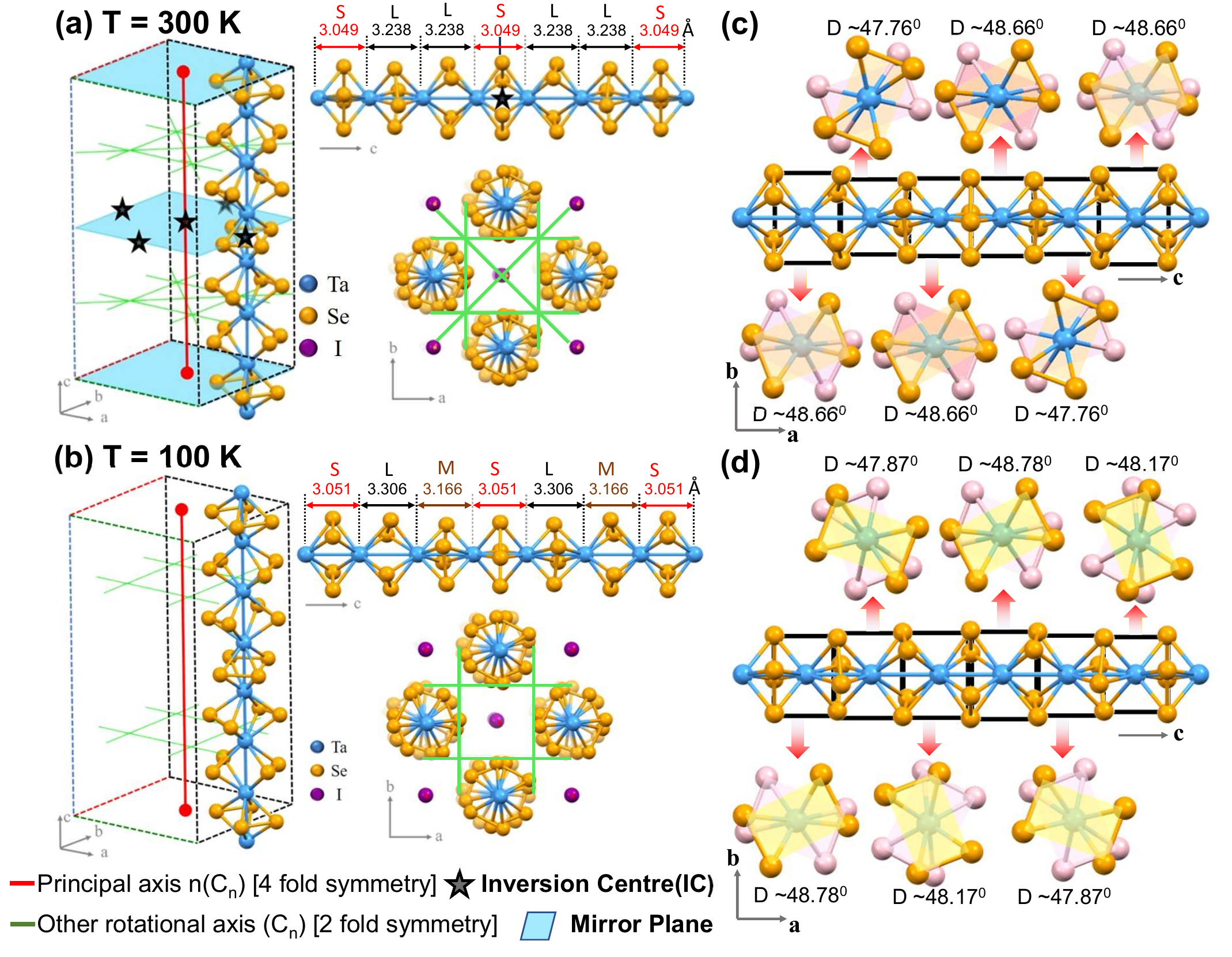}
\caption {\textbf{Single crystal XRDs: Structures and Symmetries of (TaSe$_4$)$_3$I single crystal.} Schematic of the crystal structure of (TaSe$_4$)$_3$I, at (a) $T$ = 300~K (RT), and (b) $T$ = 100~K (LT). The " Inversion Centre (IC) " points are highlighted with star symbols. The "IC"s only exist in the RT crystal structure. Similarly, mirror planes (shown in light blue color) are visible in the crystal structure at RT, whereas they are absent in the LT crystal structure. For the simplification, different inequivalent Ta sites (Ta(1) and Ta(2) in HT
and Ta(1), Ta(2), Ta(3) in LT) are shown in same color.
(c-d) The distribution of the dihedral angles in HT and LT (see text for details)}
\label{fig2:symmetry}
\end{center}
\end{figure*}

\subsection{Confirmation of "structural phase transition (SPT)":  Broken inversion symmetry}

The temperature-dependent single-crystal XRD studies were carried out to understand the structural changes across the phase transition.

\textbf{Crystal structure at $T$ =~300~K:} At room temperature (TaSe$_4$)$_3$I crystallizes in Tetragonal crystal structure (space group $P4/mnc$, no. 128). The lattice parameters of  $P4/mnc$ structure are found to be $a=b=9.4696(5)\,\text{\AA}$, $c=19.049(11)\,\text{\AA}$; $\alpha=\beta=\gamma=90^{\circ}$. The complete details of the lattice constants, structural parameters, and $R$ factors are listed in Table ~\ref{tab:widgets}. The RT crystal structure is shown in Figure~\ref{fig2:symmetry}(a). 

Key features of the RT crystal structure can be summarized as follows,
(i) Unit cell of (TaSe$_4$)$_3$I consists of parallel TaSe$_4$ chains, which are well separated from each other by I atoms as shown in Figure~\ref{fig2:symmetry}(a). The iodine atoms provide cohesion to the structure, without active involvement in the bonding. (ii) There are twelve Ta-atoms of two inequivalent classes, Ta(1) and Ta(2) in a unit cell. Each Ta atom is sandwiched between two nearly rectangular Se$_4$ units. The shorter Se-Se side of each Se$_4$ has a bond length of about 2.35-2.36 $\text{\AA}$, a distance typical of a Se$_2^{2-}$ dimer \cite{Gressier1984a}, while the longer Se-Se side is about 3.58-3.59 $\text{\AA}$. Any two adjacent Se$_4$ units make an angle around 45$^0$ to 50$^0$. (iii) The Ta-Ta bonding along the chain exhibits the sequence of $\ldots$-Short(S)--Long(L)--Long(L) $\ldots$ bonds. (iv) The distance between nearest TaSe$_4$ chains, $d_{inter}=6.677\,\text{\AA}$ and the diameter of each TaSe$_4$ chain,$d_{intra}=4.271$\,$\text{\AA}$. The distance between two opposite-faced TaSe$_4$ chains is 9.4 $\text{\AA}$. This confirms the quasi-one-dimensional nature of the crystal structure  with strong covalent bonds along the 1-D chain direction, and weak interactions between the chains\cite{Gressier1984a,Monceau2012,FavreNicolin2001}.

\begin{figure}
\begin{center}
\includegraphics[width=1\columnwidth]{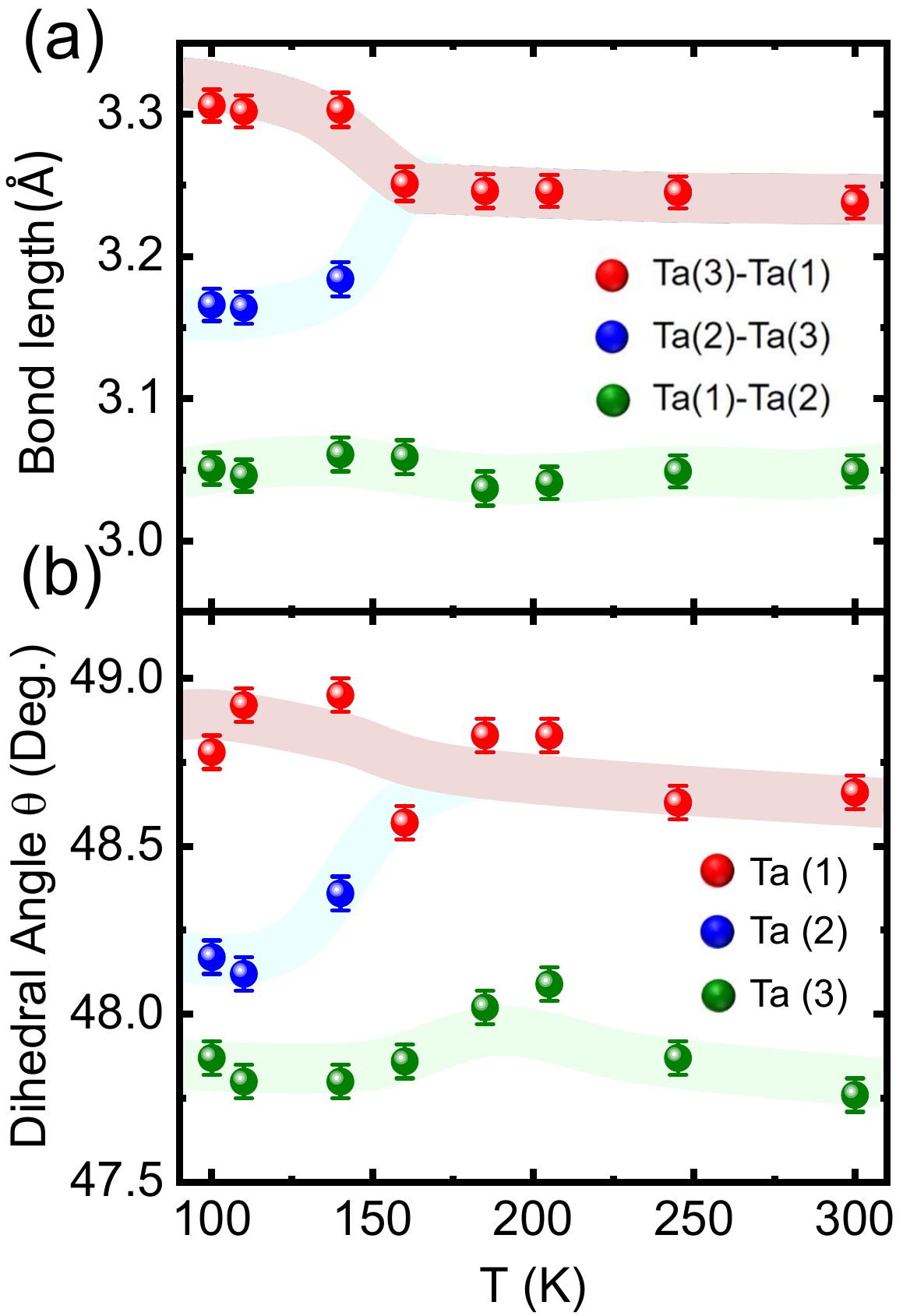}
\caption {(a) Temperature variation of Bond distance among different Ta atoms and (b) dihedral angle. The solid sphere represents the experimental data, while thick solid lines are the guide to the eye.}
\label{fig3:bonddistanceAndbondangle}
\end{center}
\end{figure}

\textbf{Crystal structure at $T$ = 100~K :} 
The SXRD experiment performed at $T$ = 100~K reveals that low-temperature crystal structure also belongs to the tetragonal symmetry but with a different space group, P$\Bar{4}$2$_1$c (no. 114). Interestingly, the space group of the low-temperature crystal structure is non-centrosymmetric, in contrast to the centrosymmetric nature of the room-temperature $P4/mnc$ space group. Although the  refined  lattice parameters of the LT phase: $a=b=9.4358(5)\,\text{\AA}$, $c=19.0464(11)\,\text{\AA}$; $\alpha=\beta=\gamma=90^{\circ}$, are found to be similar to the lattice parameters of the crystal structure at RT. The complete crystallographic information  is given in Table ~\ref{tab:widgets}. The obtained crystal structure corresponding to $T$~=~100~K is presented in Figure~\ref{fig2:symmetry}(b). 

The essential features that capture the changes between the RT and LT crystal structure are summarized below:

(i) As mentioned above, the Ta-Ta bonding sequence along a TaSe$_4$ chain at RT structure is found to be a sequence of $\ldots$-Short (S) (3.059\,\AA)-Long (L) (3.251\,\AA)-Long (L) (3.251\,\AA)- $\ldots$ bonds. The "L" bonds are formed between Ta(1) \& Ta(2) atoms and "S" bonds are between both Ta(1) \& Ta(1) and Ta(2) \& Ta(2) atoms. On the other hand the low-temperature structure (see Figure~\ref{fig2:symmetry}(b)) exhibits Ta-Ta bonding sequence of  $\ldots$-Long(L) (3.306\,\AA)-Medium(M) (3.116\,\AA)-Short(S) (3.051\,\AA)-$\ldots$ bonds. 
Compared to the HT structure, instead of two inequivalent Ta atoms, Ta(1) and Ta(2), three
inequivalent Ta classes, Ta(1), Ta(2) and Ta(3) are formed in LT.
The "S", "M", and "L" bonds are formed between Ta(1) \& Ta(2), Ta(2) \& Ta(3), and Ta(1) \& Ta(3) atom, respectively. The Ta(3) atom is thus pulled closer to Ta(2) atom so that the bond length between Ta(2) \& Ta(3) atoms is changed from long (L) to medium (M) and the Ta(3)-Ta(1) bond-length is increased slightly, compared to the HT structure (see Figure~\ref{fig_metalbondingseq}).

(ii) As discussed, the individual chain is built by stacking of TaSe$_4$ antiprisms in a screw-like arrangement. The planes formed by both sides of Se atoms are twisted with respect to each other at a certain "dihedral angle". Figure~\ref{fig2:symmetry} (c-d) shows the sequence of "dihedral angle" at the HT structure and the LT structure. As is seen and
as expected, the changes in the Ta-Ta bond sequence at the structural transition gets also reflected in dihedral angles.

(iii) Symmetry consideration yields the following major differences between the two structures:\\
(a) RT structure hosts five inversion centers (IC). Among them, four reside in the central position of each TaSe$_4$ chain, and the remaining one is at the center of the ${Tetragonal}$ unit cell (see Figure~\ref{fig2:symmetry}(a). All five inversion centers are lost in LT structure.\\
(b) The RT crystal structure also hosts mirror planes, whereas they are absent at LT crystal structure(see Figure~\ref{fig2:symmetry}(a-b)).\\
(c) Both RT and LT structures have one principal rotational axis (n($C_4$)) which has 4-fold symmetry. A careful look reveals, the presence of 2-fold rotational axis ($C_2$), which are six in RT structure and four in LT structure.\\

\begin{table*}
\caption{\label{tab:widgets}{\textbf{SXRD Refinement results for (TaSe$_4$)$_3$I at $T$ = 100~K, and 300~K.}} }
\begin{tabular}{ |p{3cm}||p{6cm}|p{6cm}| }
\hline
\multicolumn{3}{|c|}{Crystallographic Information} \\
\hline
Temperature ($T$) & 100~K & 300~K\\
\hline
Crystal system & Tetragonal & Tetragonal\\
Space Group & P$\Bar{4}$2$_1$c & $P4/mnc$\\
$a$ (\AA) & 9.4358 & 9.4696\\
$b$ (\AA) & 9.4358 & 9.4696\\
$c$ (\AA) & 19.046 & 19.049\\
$\alpha$=$\beta$=$\gamma$ ($^{\circ}$) & 90 & 90 \\
Cell Volume, (\AA$^3$) & 1695.78 & 1708.19\\
R-factor(\%) & 2.27 & 2.75\\
\hline
\multicolumn{3}{|c|}{Atomic Coordinates} \\
\hline
\multicolumn{1}{|c||}{Atom} & \multicolumn{1}{|c|}{Site\hspace{2.5em}$x$\hspace{2.5em}$y$\hspace{2.5em}$z$\hspace{2.5em}B$_{iso}$} & \multicolumn{1}{|c|}{Site\hspace{2.5em}$x$\hspace{2.5em}$y$\hspace{2.5em}$z$\hspace{2.5em}B$_{iso}$}\\
\hline
\multicolumn{1}{|c||}{Ta(1)} & \multicolumn{1}{|c|}{4d\hspace{2em}0.50\hspace{2em}1\hspace{2em}0.913\hspace{2em}0.009} & \multicolumn{1}{|c|}{8f\hspace{2em}0\hspace{2em}0.50\hspace{2em}0.58\hspace{2em}0.021}\\
\multicolumn{1}{|c||}{Ta(2)} & \multicolumn{1}{|c|}{4d\hspace{2em}0.50\hspace{2em}1\hspace{2em}0.573\hspace{2em}0.009} & \multicolumn{1}{|c|}{4d\hspace{2em}0\hspace{2em}0.50\hspace{2em}0.75\hspace{2em}0.023}\\
\multicolumn{1}{|c||}{Ta(3)} & \multicolumn{1}{|c|}{4d\hspace{2em}0.50\hspace{2em}1\hspace{2em}0.739\hspace{2em}0.009} & \multicolumn{1}{|c|}{-\hspace{3em}-\hspace{3em}-\hspace{3em}-\hspace{3em}-}\\
\multicolumn{1}{|c||}{Se(1)} & \multicolumn{1}{|c|}{8e\hspace{1.5em}0.547\hspace{1.5em}0.785\hspace{1.5em}0.820\hspace{1.5em}0.011} & \multicolumn{1}{|c|}{16i\hspace{1.5em}0.186\hspace{1.5em}0.375\hspace{1.5em}0.66\hspace{1.5em}0.024}\\
\multicolumn{1}{|c||}{Se(2)} & \multicolumn{1}{|c|}{8e\hspace{1.5em}0.313\hspace{1.5em}0.876\hspace{1.5em}0.832\hspace{1.5em}0.011} & \multicolumn{1}{|c|}{16i\hspace{1.5em}-0.048\hspace{1.5em}0.28\hspace{1.5em}0.672\hspace{1.5em}0.025}\\
\multicolumn{1}{|c||}{Se(3)} & \multicolumn{1}{|c|}{8e\hspace{1.5em}0.283\hspace{1.5em}1.049\hspace{1.5em}0.663\hspace{1.5em}0.011} & \multicolumn{1}{|c|}{8h\hspace{1.6em}0.021\hspace{1.6em}0.275\hspace{1.6em}0.50\hspace{1.6em}0.026}\\
\multicolumn{1}{|c||}{Se(4)} & \multicolumn{1}{|c|}{8e\hspace{1.5em}0.372\hspace{1.5em}0.814\hspace{1.5em}0.653\hspace{1.5em}0.011} & \multicolumn{1}{|c|}{8h\hspace{1.5em}-0.199\hspace{1.5em}0.391\hspace{1.5em}0.50\hspace{1.5em}0.027}\\
\multicolumn{1}{|c||}{Se(5)} & \multicolumn{1}{|c|}{8e\hspace{1.5em}0.610\hspace{1.5em}0.800\hspace{1.5em}0.493\hspace{1.5em}0.012} & \multicolumn{1}{|c|}{-\hspace{3em}-\hspace{3em}-\hspace{3em}-\hspace{3em}-}\\
\multicolumn{1}{|c||}{Se(6)} & \multicolumn{1}{|c|}{8e\hspace{1.5em}0.725\hspace{1.5em}1.022\hspace{1.5em}0.492\hspace{1.5em}0.011} & \multicolumn{1}{|c|}{-\hspace{3em}-\hspace{3em}-\hspace{3em}-\hspace{3em}-}\\
\multicolumn{1}{|c||}{I} & \multicolumn{1}{|c|}{4c\hspace{1.8em}0.50\hspace{1.8em}0.50\hspace{1.8em}0.626\hspace{1.8em}0.022} & \multicolumn{1}{|c|}{4e\hspace{1.8em}0.50\hspace{1.8em}0.50\hspace{1.8em}0.627\hspace{1.8em}0.056}\\
\hline
\end{tabular}
\end{table*}

\textbf{Temperature-dependent crystal structures:} The temperature-dependent SXRD study confirms that there is an inversion symmetry breaking accompanied by a structural transition from HT centrosymmetric phase to  a LT noncentrosymmetric phase. Figure~\ref{fig3:bonddistanceAndbondangle}(a-b) shows the temperature variation of bond distance between different Ta-atoms and dihedral angles along the chains revealing the structural transition at around 145~K. The primary observations are:

(a) With decreasing temperature, the bond distance between Ta(1) \& Ta(2) (; S) sites remains similar throughout the entire temperature scan, whereas the distance between Ta(3) \& Ta(1) (; L) and Ta(2) \& Ta(3) (; L) follows a similar path down to 160 K. Below $T_S$, a slight shift of Ta(3) atom along the chain changes the bond-length between Ta(2) \& Ta(3) atoms from long (L) to medium (M), as a result, the Ta(3)-Ta(1) bond length increased slightly to keep the sum of bond lengths constant (see Figure~\ref{fig3:bonddistanceAndbondangle}(a)). These rearrangements of Ta-atoms lead to the breaking of inversion symmetry in the crystal structure. 

(b) The dihedral angle of Ta(3) centered antiprism remains almost unchanged throughout the entire temperature variation while the dihedral angle of Ta(1) and Ta(2) centered antiprisms  behave in a similar  manner down to 160~K. Below 160K, the  Ta(1) centered dihedral angle is  increases and Ta(2) centered dihedral angle decreases with further lowering of the temperature (see Figure~\ref{fig3:bonddistanceAndbondangle}(b)).\\

The temperature-dependent SXRD data (see Figure~\ref{fig_metalbondingseq} for more details on the  structure) thus corroborate a unique centrosymmetric to noncentrosymmetric structural phase transition belonging to the same tetragonal symmetry in (TaSe$_4$)$_3$I, at around $T_S$~$\sim$145~K.
The presence of additional Raman modes (Figure~\ref{fig1 raman and cp}) at LT is further verification of this transition, which is attributed to the absence of inversion centers.
Samples with no inversion center have vibrational modes that are both Raman and infrared active, whereas those with an inversion center have zone-center lattice vibrations with even or odd parity \cite{Kakkar2015,Bernath2016}. Hence no normal modes can be both Infrared and Raman active. The odd-parity vibration modes are Raman inactive due to selection rules, resulting in fewer Raman-active modes \cite{Kakkar2015,Bernath2016,Bera2023Raman}.  To determine the number of Raman-active modes for each phase, we conducted group theory analysis. Since both phases contain 64 atoms, the total number of mechanical vibrational modes is 192, of which 3 are acoustic and 189 are optical. The HT centrosymmetric phase has 64 infrared active modes and 83 modes that are both Raman and infrared active. However, with decreasing temperature below the phase transition, the 64 Raman-inactive modes of the centrosymmetric phase become Raman-active in the noncentrosymmetric phase.  This results in a larger number of Raman-active modes in the  LT noncentrosymmetric phase, as seen in Raman spectra presented in Figure~\ref{fig1 raman and cp}\cite{Bera2023Raman}.

\subsection{Theoretical calculation:}

\begin{figure*}
\begin{center}
\includegraphics[width=2\columnwidth]{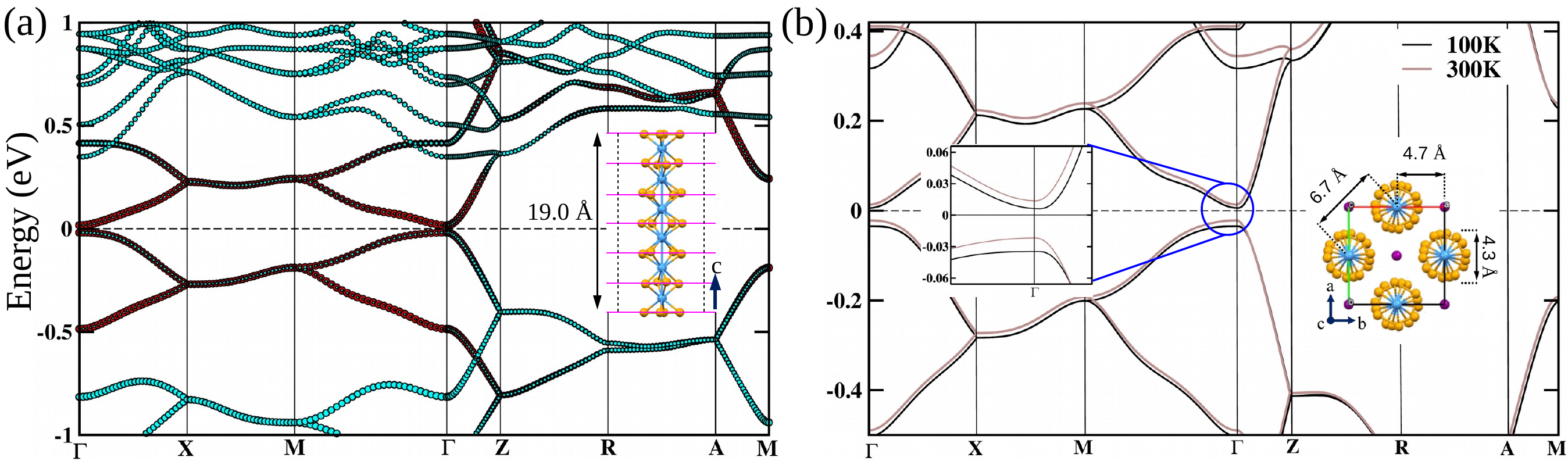}
\caption{(a) The band structure of (TaSe$_4$)$_3$I at HT (T~=~300~K) crystal structure, plotted along the high symmetry k-points, $\Gamma$=(0,0,0)-
X($\pi/a$,0,0)-M($\pi/a$,$\pi/a$,0)-$\Gamma$-Z(0,0,$\pi/c$)-R($\pi/a$,0,$\pi/c$)-A($\pi/a$,$\pi/a$,$\pi/c$)-M($\pi/a$,$\pi/a$,0). The zero of the energy is set at the
Fermi energy. The fatness of the bands indicates the orbital characters (red for Ta-$d$, cyan for Se-$p$. The inset shows the TaSe$_4$ chain running along the crystallographic
$c$-axis. (b) Comparison of the band structure for LT (100~K) crystal structure with that at RT (300~K). The change in the band gap between the two band structure is shown in the zoomed-in plot. The inset shows the TaSe$_4$ chains,
separated by I atoms, in (TaSe$_4$)$_3$I structure, projected in the $ab$ plane.}
\label{BandStructure}
\end{center}
\end{figure*}

In order to uncover the origin of the structural transition, we carried out first-principles electronic structure calculations of (TaSe$_4$)$_3$I at high-temperature (T=300~K) (HT) and low-temperature (T=100~K) (LT) crystal structures.
The left panel of Figure~\ref{BandStructure}, shows the band structure of (TaSe$_4$)$_3$I at 300~K, plotted along
the high symmetry k-points of the tetragonal BZ and projected
to Ta-$d$ and Se-$p$ states. As mentioned above, each Se$_4$ rectangle is made up of two Se$_2^{2-}$ dimers. Each Ta atom in TaSe$_4$ chain in the rectangular antiprism Se
environment of Se atoms is thus in nominal 4+ or $d^1$ charge state. The electron withdrawal effect of iodine, present in between the chains, reduces the nominal valence
of Ta from 4+, and makes it fractional to 4.333+ or d$^{2/3}$. As a linear chain
with fractional occupancy, $f <$ 1, leads to Peierls distortion, the T=300~K structure
shows trimerization with two inequivalent Ta atoms (Ta(1) and Ta(2)) and two long (L)
and one short (S) Ta-Ta bond of sequence $\ldots$ L-L-S $\ldots$, as found in the
experimentally measured structure. The Peierls distortion opens up a gap at the Fermi level,
the magnitude of which depends on the strength of the trimerization. Our calculated
electronic structure shows a tiny gap of 0.02 eV, making the calculated electronic
structure almost semimetal. From the projected orbital characters of the bands, it is
seen that the low-energy band structure is dominated by Ta-$d$, with some admixture
from Se-$p$. The rectangular antiprism coordination of Se atoms surrounding Ta, puts
Ta $3z^2$-$r^{2}$ as the lowest energy state, which is about half-filled for Ta(1) and about empty for Ta(2). There are 12 Ta atoms in the unit cells, with 8 Ta(1) and 4 Ta(2), giving rise
to eight Ta(1)-$3z^2$-$r^{2}$ dominated bands and four Ta(2)-$3z^2$-$r^{2}$ bands. The overlap
between Ta(1)-$3z^2$-$r^{2}$ and Ta(1)/Ta(2)-$3z^2$-$r^{2}$, splits the eight Ta(1)-$3z^2$-$r^{2}$ dominated bands into a group of four occupied and four unoccupied bands, while the four Ta(2)-$3z^2$-$r^{2}$ bands remain more or less unoccupied. 
We find the bands are almost flat along X-M, {\it i.e} going along the $b$-direction, where two chains are separated by a distance of about 9.5{~\AA} with intervening I$^{-1}$ ions, which remain isolated (cf insets). The occupied 
bands along M-$\Gamma$ {\it i.e.} moving along the diagonal in $ab$-plane further splits due to Ta-Ta interaction via the intervening Se atoms of two neighboring chains (cf insets). The trimerization along the chain direction, $c$, causes further
band folding, as observed in bands structure along $\Gamma$-Z direction. The basic features of the electronic structure of 100~K crystal structure, remain essential same as that of the high-temperature crystal structure (cf right panel of Figure \ref{BandStructure} for comparison). Twelve Ta atoms in the unit cell are now split into three classes, 4 Ta(1),
4 Ta(2), and 4 Ta(3). The occupancies of Ta(1) and Ta(2) are found to be similar, being close
to half-filled, while that of Ta(3) is found to be different, being more empty.
We find that the band gap shows a tiny increase in the low-temperature phase 
from 0.02 eV to 0.03 eV, ensuring increased stability of the LT phase
over the HT phase. It is worth mentioning here that the existence of the tiny but non-zero gap in the electronic structure of both HT and LT phases, as well as the enhancement of band gap value in LT phase over the HT phase have been confirmed by calculations in three different methods, plane-wave, full potential LAPW as well as LMTO. This makes the compound semi-metal, as seen in the experiment.

\begin{figure}
\begin{center}
\includegraphics[width=1.0\columnwidth]{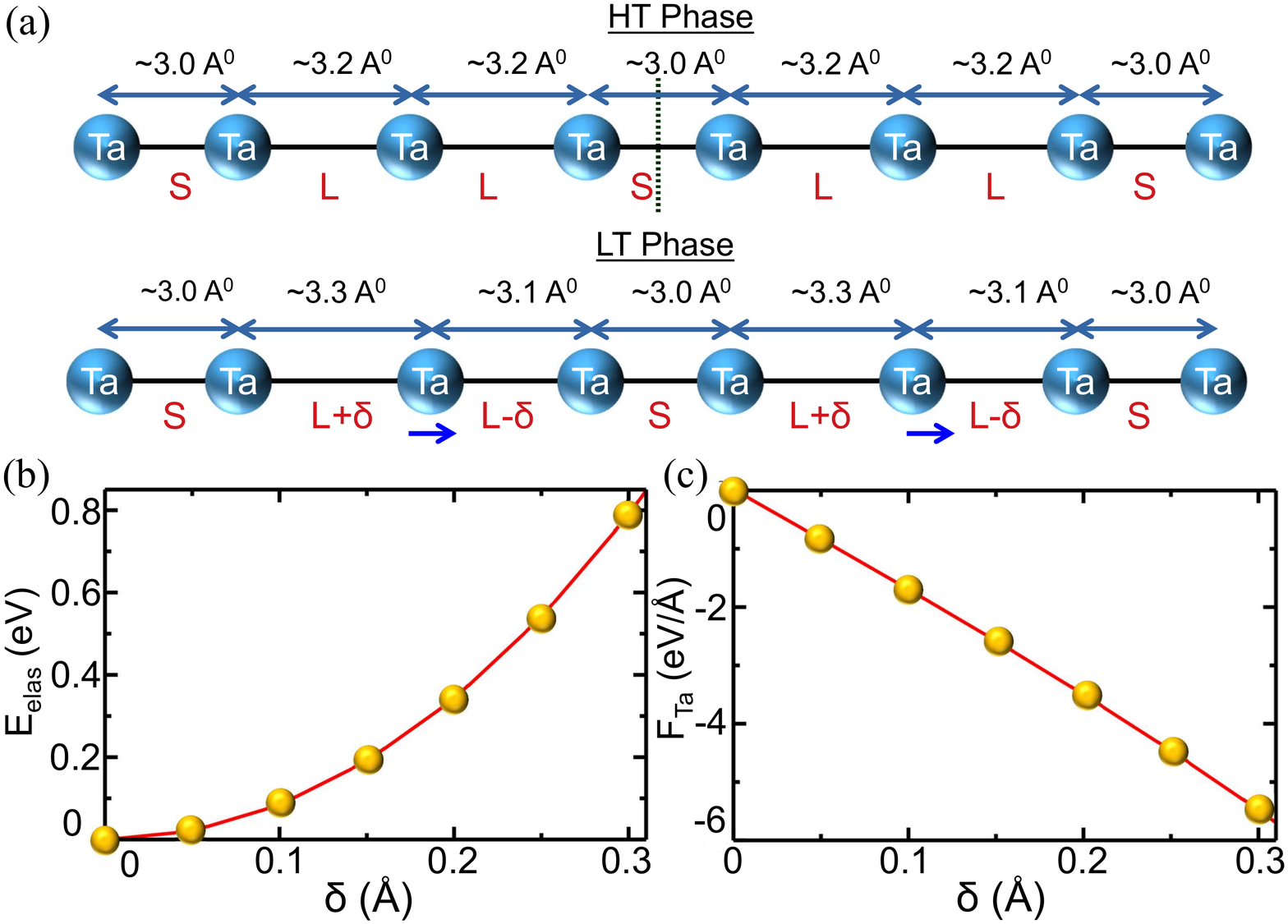}
\caption{(a) The schematic diagram of the distribution of Ta-Ta bond lengths in the  TaSe$_4$ chain in HT and LT crystal structure. The center of inversion (marked as a dotted line) in the HT structure, gets broken in the LT structure due to the distortion $\delta$. Elastic energy (b) and force acting on Ta ions (c) plotted as a function of off-centric displacement ($\delta$), obtained from DFT calculations.}
\label{ElasticEnergy}
\end{center}
\end{figure}

In order to understand the origin of the HT-LT phase transition, we note that the low-temperature crystal structure, instead of $\ldots$L-L-S-L-L$\ldots$ sequence of Ta-Ta
bonds in the chain of the HT phase adopt the $\ldots$L-M-S-L-M$\ldots$
sequence (see Figure~\ref{ElasticEnergy} schematic diagram). Thus the inversion symmetry at the center of "S" bond at HT phase, breaks down at LT phase, making the transition akin to off-centric movement in ferroelectrics, with centrosymmetric HT and noncentrosymmetric LT crystal structures. The off-centric movement in ferroelectrics has been argued to arise from the gain in hybridization
energy due to the distortion, balanced by the loss in elastic energy. In order to estimate the hybridization energy gain, we considered a Ta-$3z^2$-$r^2$ only model. We note that between HT and LT structure, the "S" bond length remains more or less, unchanged while the "L" bond at HT, splits into $\approx$ "L+$\delta$" and $\approx$ "L-$\delta$. We used the NMTO-downfolding method\cite{nmto} to construct the Ta-$3z^2$-$r^2$ only model of the HT electronic structure, by integrating all the degrees of freedom, except Ta-$3z^2$-$r^2$ starting from the density functional theory (DFT) band structure of the HT phase. The tight-binding Hamiltonian, constructed in the basis of downfolded  Ta-$3z^2$-$r^2$ Wannier function provided the estimate of hopping, $t_{dd}$ corresponding to "L"
bond. The estimate of hopping, corresponding to LT phase, was obtained by distance ($l$)-dependent scaling relation, $t_{dd}$ $\sim$ $1/{l^\beta}$ where $\beta$ = ${\partial ln t_{dd}}/ { \partial ln l}$. From the estimates of $t_{dd}$ corresponding to "L" and "S" bonds of the HT phase, $\beta$ turned out to be $\sim$ 2. The kinetic energy gain due to
additional off-centric distortion of the chain at LT over that in HT is thus given by,
\[
\Delta E_{KE} = -2t_{dd}^2 (L+\delta) - 2t_{dd}^{2} (L-\delta) + 4 t_{dd}^{2}
\]
The elastic energy cost due to two "L" bonds to "L+$\delta$" and "L-$\delta$ is given by
\[
\Delta E_{elas} = \kappa \times \delta^2
\]
where $\kappa$ is the stiffness constant. We estimated the stiffness
constant from total energy calculations carried out on plane-wave basis upon
varying values of $\delta$. This is shown in Figure~\ref{ElasticEnergy}. The stiffness constant can then be extracted by fitting the evolution of the energy as a function of $\delta$ with the function $f(\delta)$ = 
$\kappa\times\delta^2$/2. We also adopted an alternative procedure of extracting the stiffness constant from the force acting on the Ta ions and fitting the data points 
with the function $f^{'}(\delta)$. The two procedures gave rise to a similar value of $\kappa$ 16.8 eV/\AA$^2$ from total energy and 17.0 eV/\AA$^{2}$ from force estimate. Plugging in the values of t$_{dd}$, $\delta$, $\kappa$, lead to $\Delta E_{KE}$ of $\sim$ -0.03 eV, and $\Delta E_{elas}$ $\sim$ 0.01 eV, making the distorted LT structure stabler due to net hybridization energy gain. It is to be noted that the above simplistic analysis does not take into account the additional source of distortion, namely the change in the dihedral angle between the $Se-$ rectangles. We also add that the above theoretical analysis is restricted to only one chain and the loss of the inversion center in the chain. Following the above analysis, one would expect (TaSe$_4$)$_3$I to exhibit finite polarization. However, the experimentally determined P$\Bar{4}2_1c$ (114) space group though non-centrosymmetric, is non-polar as the corresponding point group $\Bar{4}$2m is non-polar. This is caused by the additional loss of inversion in the center of the tetragonal cell in LT. The Ta-Ta bonding sequence in a given chain in LT structure is $\ldots$-L-M-S-$\ldots$, while it changes to $\ldots$-M-L-S-$\ldots$ in the neighboring chain. As checked explicitly, for identical bonding sequences between the chains, as in HT structure, the LT space group would have been P4nc, with a polar point group, 4mm, supporting the theoretical conjecture.

\section{Conclusion}

In summary, employing detailed analysis of the single crystal XRD data, we have characterized the structural transition  at around $T_s \sim 145~K$ in (TaSe$_4$)$_3$I, evidenced in resistivity, specific heat, and Raman scattering experiments. The structural transition turned out to be a symmetry-lowering transition of the trimerized Ta chains from the L-L-S sequence of Ta-Ta bonds to L-M-S Ta-Ta bonds, thereby breaking the center of inversion at the short Ta-Ta bond. This causes a transition between a  high-temperature (T$>T_s$) centrosymmetric structure and a low-temperature (T$<T_s$) noncentrosymmetric structure belonging to the same tetragonal symmetry. The origin of this structural phase transition is examined in terms of first-principles DFT calculations and tight-binding modeling. We find the high and low-temperature structures to be both semi-metallic consistent with the recent report. The breakdown of the inversion symmetry in the low-temperature structure is found to be caused by the off-centric movement of Ta atoms in the Ta chain structure, resulting in a gain in the hybridization energy. This is balanced by elastic energy loss in terms of expansion and contraction of the bonds. First-principles derived estimates of the hybridization energy gain and elastic energy loss suggest that the hybridization energy gain wins over the elastic energy loss, thereby stabilizing the low-temperature noncentrosymmetric structure. It is to be noted that this noncentrosymmetric structure remains non-polar due to additional lifting of the inversion center between the chains. Our work opens up new opportunities for exploring novel quantum phenomena in quasi-1D materials with broken inversion symmetry and could lead to the development of new quasi-1D materials with unique properties for potential applications.

\section{Acknowledgement}
This work was supported by the (i) 'Department of Science and Technology, Government of India (Grant No. SRG/2019/000674 and EMR/2016/005437), and (ii) Collaborative Research Scheme (CRS) project proposal(20opensRS/59/58/760). A.B. thanks CSIR Govt. of India for Research Fellowship with Grant No. 09/080(1109)/2019-EMR-I. A.B. acknowledges the experimental facilities for sample growth, procured using financial support from DST-SERB grant nos. ECR/2017/002 037. T.S-D acknowledges J.C.Bose National Fellowship (grant no. JCB/2020/000004) for funding. M.M. and A.B. would like to thank Dr. Atindra Nath Pal, Prof. Goutam Sheet, and Pof. Prabhat Mandal for the technical help and  discussion. A.B. would like to acknowledge  Dr. Mainak Palit for his technical help with Raman spectroscopy. 

\appendix
\renewcommand{\thefigure}{A\arabic{figure}}
\renewcommand{\thesection}{A\arabic{section}}
\setcounter{figure}{0}
\setcounter{section}{0}

\begin{figure*}
\begin{center}
		\includegraphics[width=2\columnwidth]{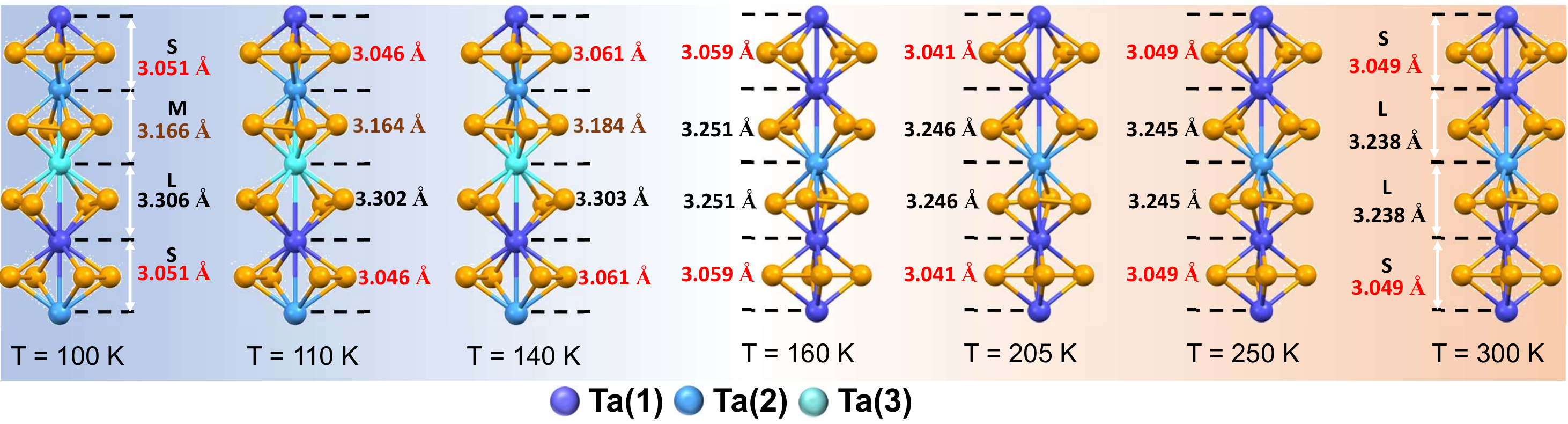}
		\caption {\textbf{$Ta$ metal bonding sequence in n-TSI:} Metal bonding sequence, obtained from SXRD data, along each TaSe$_4$ chain at $T$ = 100~K, 110~K, 140~K, 160~K, 205~K, 250~K, and 300~K.}
		\label{fig_metalbondingseq}
	\end{center}
\end{figure*}

\section{Evolution of structure across transition}

Fig.~\ref{fig_metalbondingseq} represent the thermal evolution of $Ta$ metal bonding sequence of a TaSe$_4$ chain obtained from single-crystal X-ray diffraction (SXRD) data at various temperatures ranging from 100~K to 300~K. At each temperature, the sequence of metal-metal bonds along the TaSe$_4$ chain was determined from SXRD data. The results reveal the temperature-dependent nature of the TaSe$_4$ chain structure as the temperature increases across the transition temperature, $T_S$.

%

\end{document}